\documentclass[runningheads]{llncs}
\usepackage{graphicx}

\usepackage[backend=bibtex, style=alphabetic, sorting=ynt]{biblatex}

\usepackage{graphicx}

\bibliography{SBLivingLab.bib}

\begin{document}
\title{Living Innovation Lab: a Human Centric Computing toward Healthy Living\thanks{}}
%
%

\author{Swati BANERJEE\inst{1}} 
	

%
\authorrunning{Swati BANERJEE et al.}
%

\institute{LINEACT EA:7527, CESI, Dijon \\
\url{https://lineact.cesi.fr/en/} \\
\email{\{swatibanerjee\}@ieee.org}}

\maketitle              
\begin{abstract}
Living Lab is an umbrella term used for referring to a methodology of user-centric innovation in real-life environments within a wider network of relevant stake holders. Real-life environment refers to living houses and hospitals inter wined and connected together in a way which promotes direct usability of research by the end users. It primarily consists of three stages, Design thinking to actual Conceptualisation, Evaluation and Prototyping and Final product prototyping to commercialisation.  
The increasing demand of cutting age healthcare system is in itself a challenge and requires user involvement to mobilise knowledge to build a patient centered and knowledge-based economy. Innovations are constantly needed to reduce the problematic barriers to efficient knowledge exchange and improve collaborative problem solving. Living Innovation Lab, as open knowledge system, have immense potential to address these gaps that are underexplored in the healthcare system.

\keywords{Living Lab \and Human Centric Computing \and Design and Conceptualizing \and Evaluation and Prototyping.}
\end{abstract}
\section{Introduction - Living Lab Concept}

Living Lab - a term that was first used at the Massachusetts Institute of Technology (MIT). It proposes a research methodology that is not only user-centred but “carried by the users” permitting the formulation, prototyping and validation of complex solutions in a multifaceted real-life environment\cite{businessModelSmartCity}.
It is estimated that as much as 85\% of the problems with new products and analysis pipelines originate from a poor design process \cite{RajivGrover}; enterprises carrying out product development are under constant pressure to improve their design processes to stay competitive in an ever demanding competitive market. Simultaneously, the necessity for a more cost-effective and quick development of products, services and applications in the majority of these businesses remains inevitable. However, a significant number of well-developed technologies are lacking a sufficiently marketable application or service and only 15\% of product development time is invested in products which reach the real market to actual users. 

In order to achieve faster return on investment, developments are often based solely on technological possibilities, not on the actual needs of customers. A result of this practice, only 18\% of the innovations brought into the market prove sustainably successful \cite{Cappellin2009}.
It follows that in order to reduce risks in the product development process, customers and other stakeholders need to be more directly integrated. One concept for such integration is that of Living Labs. It is a systemic innovation approach in which all stakeholders in a product, service or application participate directly in the development process. It refers to a research and development (R\&D) methodology in which innovations are created and validated collaboratively in multi-contextual, empirical real-world environments. The individual is in the focus in his or her role of as, for example, a citizen, consumer or worker. In Living Labs, collaborative Information and Communication Technologies (ICT) provide the basis for targeted customer-centred development. Given new possibilities to participate in emerging value networks, he or she can act as much as a producer than as a consumer.

\section{Components of a Living Lab}

\subsection{Design Thinking to Actual Conceptualization}

Living Lab is an umbrella concept used for a diverse set of innovation milieus emerging all over the world. Even though they differ in many ways, both in focus and approach, there also exits a few common denominators pulling them all together \cite{LivingLabsAndUserEngagement}.
In order to create a joint venture, coordinate activities, and share learning experiences, a European Network of Living Labs has developed. The aim of the network is to offer a gradually growing set of networked services to support the "Innovation Lifecycle" for all actors in the system: end-users, SMEs, corporations, the public sector, and academia \cite{Schumacher2007LivingL}. 

Figure 1, shows some of the main components of the Living Lab. The development flow is divided into 3 main sub-blocks, Conceptualization--Evaluation--Product Prototype Delivery 
\\The first part deals with from Evaluation of concepts, to generating needs and Planning. The second part is the Design prototyping and evaluation of the concept under study and finally the third and the final step is the delivery of the product prototype for user testing and hence commercialisation.  

One of the key components of the Living Lab is the real-life environment test. Testing a data service pipeline or a prototype product in a realistic looking condition and test-bench environment. This allows for a long term technical solutions to be drawn up that will be sustainable and full of value both from the developer and user aspect. 

\begin{figure}[ht]
\includegraphics[width=\textwidth]{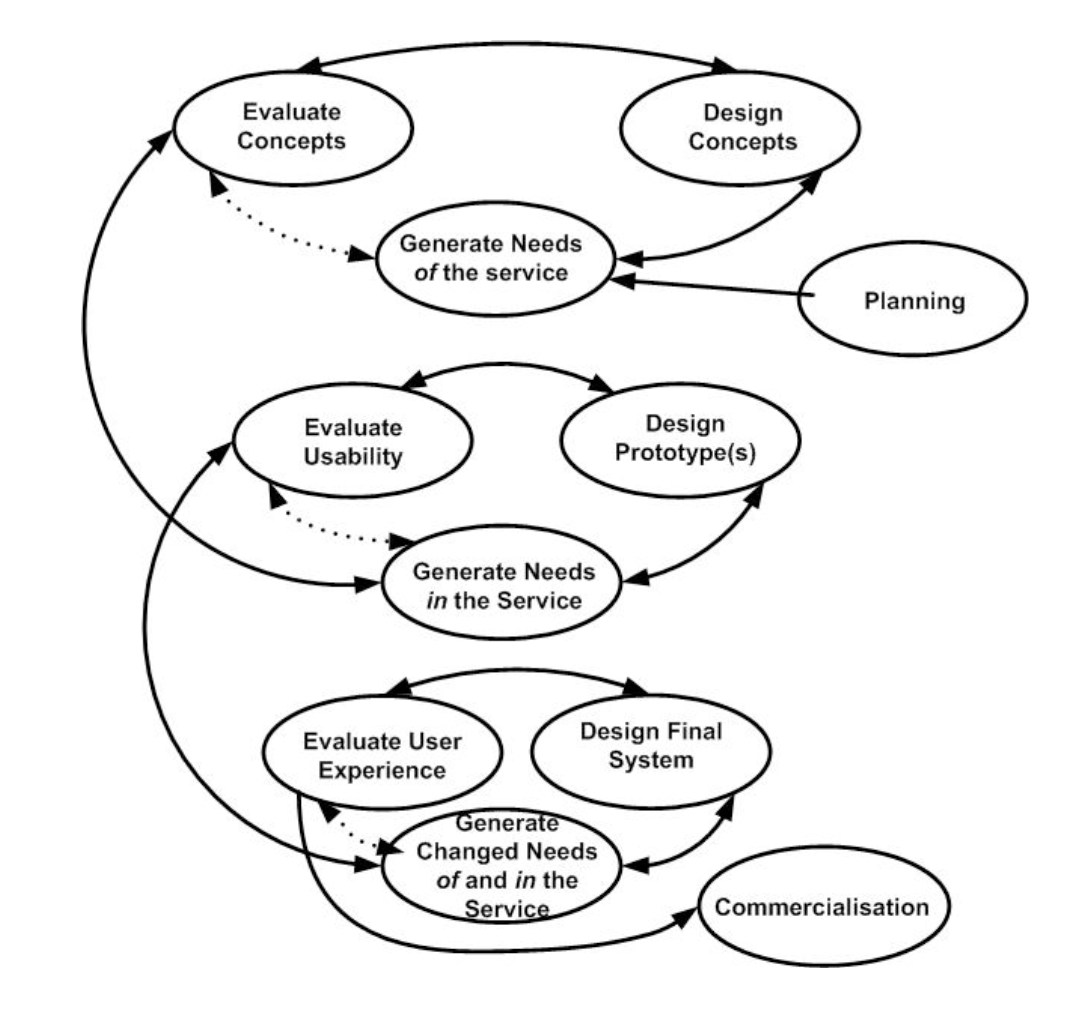}
\caption{Components of A Living Lab}
\label{fig:figure1}
\end{figure}

\subsection{Evaluation and Prototyping}
The other pillar of the Living Lab is the principle of co-creation. In a Living Lab, users are not simply the end-users but a community of people interacting with the final product or services. This concept includes not only end users but also upstream users and secondary users. At this step the actual evaluation of the device starts along with building the analysis framework, which can be based on virtual to actual prototyping to support the comprehensive evaluation of the system under consideration. This type of framework enables functional and timely verification of the prototype under consideration. Performance and reliability analysis is another important aspect to be considered while reducing the computational and economic cost. Additionally design space exploration both in terms of hardware and software of the over all system takes precedence.        

\subsection{Final Product Prototyping to Commercialisation}

From a business perspective, it is useful to perceive Living Labs as providing a set of distinct services to their customers. 
The product which comprises of sensors and the associated data manipulation pipeline is being developed and benchmark-ed, needs to be properly packaged leading to commercialising. In this regard, the customers are for example, SMEs, industry, research or public and civic organisations along with the actual people using and evaluation the product. At this stage the services offered by the living lab can be looked upon as a core set of co-creation services supported by services for both integration and data preparation.

\section{Application to health care facility}
There exists many living labs that are currently active in healthcare innovation for assisted living. The focus of such living lab is the acceptance and use of Internet of Things(IOT) based cyber physical systems (sensors) to support both elderly people and people with chronic condition, which is also known as ambient assisted living.
The aim is to build and enhance living CAPABILITY of the people under consideration which allows them to live independently at home longer, through the use of automated smart home devices. Inclusion of eHealth based apps are also integrated in the environment. The tasks involved are communication, protection through observation and safety using sensors, smart camera based monitoring systems and safety alarms. There remains a possibility to increase the in-house participation in fitness activities and safe sports. Integration of eHealth allows the user to be continuously monitored by keeping a tab on their vital health statistics like, blood pressure monitoring, blood sugar level, heart rate monitoring devise etc., from a distance.

\section{Challenges}
As stated in the introduction section, new proven and efficient strategies need to be adopted to both improve the design process of new products, pipelines and services. Simultaneously achieving a better product development time and successfully marketing the products packed with the associated pipeline should be the final goal.

Some other challenges associated to this issue is integration of infrastructure, the alignment of methodological aspects and convergence of policy factors.
With respect to methodological aspects, the key factor of differentiation of a Living Lab in comparison to other forms of co-operation, such as cluster of virtual breeding environments, is the direct involvement of users. So, the real challenge may lie in involving and convincing users in a sociological sense, which cab be only done effectively by taking into account the micro-context of their everyday lives.

Significant research effort must be allocated to the development of both methodologies and supporting tools which enable such integration in the most un-obstructive way possible. Effective and timely interaction with the end users without violating ethics is another challenge in terms of user acceptability of the experiments to be performed on the platform.

Another challenge is to bring co-creating users together and build a technological platform such as a Collaborative Working Environment (CWE) for the support co-creating process must receive a higher task priority. Such an efficient environment will be required to support all the creative stake holders.

This bring us to the most important part of the discussion, which is the protection and implementation of the IPR (Intellectual Property Right).
As private persons become a source of ideas and innovations, an appropriate rewarding and incentive mechanism needs to be put in place which simultaneously secures pay-back to all the actors involved whilst adopting fair and suitable mechanisms for the handling of IPR and other ethical issues. According to \cite{Goodpractices}, research is furthermore required in order to create comprehensive models and methods by which experiments can be analysed and values measured to protect infringements.

%
%
%

 \printbibliography

\end{document}